\documentclass[amsfonts, amssymb, amsmath, showkeys, nofootinbib, preprint,floatfix, prl, superscriptaddress]{revtex4-2}

\usepackage[colorinlistoftodos, color=green!40,prependcaption]{todonotes}
\usepackage{graphicx}% Include figure files
\usepackage{dcolumn}% Align table columns on decimal point
\usepackage{bm}% bold math
\usepackage{physics}
\usepackage{float}
\usepackage{xcolor}
\usepackage{amsmath}
\usepackage{amssymb}
\usepackage{units}
\usepackage[english]{babel}
\usepackage{placeins}
\usepackage{lineno}

\usepackage{lipsum}

\begin{document}

\title{Universal spin wavepacket transport in van der Waals antiferromagnets}

\author{Yue Sun}
\affiliation {Department of Physics, University of California, Berkeley, California 94720, USA}
\affiliation {Materials Science Division, Lawrence Berkeley National Laboratory, Berkeley, California 94720, USA}

\author{Fanhao Meng}
\affiliation {Materials Science Division, Lawrence Berkeley National Laboratory, Berkeley, California 94720, USA}
\affiliation {Department of Materials Science and Engineering, University of California, Berkeley, California 94720, USA}

\author{Changmin Lee}
\affiliation {Materials Science Division, Lawrence Berkeley National Laboratory, Berkeley, California 94720, USA}
\affiliation {Department of Physics, Hanyang University, Seoul 04763, Republic of Korea}

\author{Aljoscha Soll}
\affiliation {Department of Inorganic Chemistry, University of Chemistry and Technology Prague, Technická 5, 166 28 Prague 6, Czech Republic}

\author{Hongrui Zhang}
\affiliation {Department of Materials Science and Engineering, University of California, Berkeley, California 94720, USA}

\author{Ramamoorthy Ramesh}
\affiliation {Department of Physics, University of California, Berkeley, California 94720, USA}
\affiliation {Materials Science Division, Lawrence Berkeley National Laboratory, Berkeley, California 94720, USA}
\affiliation {Department of Materials Science and Engineering, University of California, Berkeley, California 94720, USA}

\author{Jie Yao}
\affiliation {Materials Science Division, Lawrence Berkeley National Laboratory, Berkeley, California 94720, USA}
\affiliation {Department of Materials Science and Engineering, University of California, Berkeley, California 94720, USA}

\author{Zdeněk Sofer}
\affiliation {Department of Inorganic Chemistry, University of Chemistry and Technology Prague, Technická 5, 166 28 Prague 6, Czech Republic}

\author{Joseph Orenstein}
\affiliation {Department of Physics, University of California, Berkeley, California 94720, USA}
\affiliation {Materials Science Division, Lawrence Berkeley National Laboratory, Berkeley, California 94720, USA}

\begin{abstract}
Antiferromagnets (AFMs) are promising platforms for the transmission of quantum information via magnons (the quanta of spin waves), offering advantages over ferromagnets with regard to dissipation, speed of response, and immunity to external fields \cite{gibertiniMagnetic2DMaterials2019a,gongTwodimensionalMagneticCrystals2019a,makProbingControllingMagnetic2019a}.  Recently, it was shown that in the insulating van der Waals (vdW) semiconductor, CrSBr, strong spin-exciton coupling enables readout of magnon density and propagation using photons of visible light \cite{baeExcitoncoupledCoherentMagnons2022}. This exciting observation came with a puzzle:  photogenerated magnons were observed to propagate 10$^3$ times faster than the velocity inferred from neutron scattering \cite{scheieSpinWavesMagnetic2022}, leading to a conjecture that spin wavepackets are carried along by coupling to much faster elastic modes. Here we show, through a combination of theory and experiment, that the propagation mechanism is, instead, coupling within the magnetic degrees of freedom through long range dipole-dipole coupling. This mechanism is an inevitable consequence of Maxwell’s equations, and as such, will dominate the propagation of spin at long wavelengths in the entire class of vdW magnets currently under intense investigation.  Moreover, identifying the mechanism of spin propagation provides a set of optimization rules, as well as caveats, that are essential for any future applications of these promising systems. 
\end{abstract}

\maketitle

\section{Introduction}
Exploiting the electron's spin degree of freedom is one of the central goals of quantum information science. A promising direction is coupling spin to charge and mechanical degrees of freedom to provide interconnections in hybrid quantum systems \cite{lachance-quirionEntanglementbasedSingleshotDetection2020}. To this end, it is essential to understand and control the generation, propagation, and detection of spin information. Recent progress in magnetically ordered systems has shown the promise of using spin waves -- collective excitations of the electron spins -- to transport information over large distances \cite{pirroAdvancesCoherentMagnonics2021,cornelissenLongdistanceTransportMagnon2015,lebrunTunableLongdistanceSpin2018,lebrunLongdistanceSpintransportMorin2020,hanBirefringencelikeSpinTransport2020,weiGiantMagnonSpin2022}. Antiferromagnets (AFMs) stand at the frontier of such research, given their promise of rapid response times and insensitivity to stray magnetic fields \cite{jungwirthAntiferromagneticSpintronics2016,baltzAntiferromagneticSpintronics2018}. 

Increasingly, attention has focused on quasi-two dimensional (2D) AFMs in which planar ferromagnetic order alternates in direction from layer to layer \cite{xingMagnonTransportQuasiTwoDimensional2019}. Of particular interest are those in which magnetocrystalline anisotropy favors alignment of spin parallel to the layers \cite{lebrunTunableLongdistanceSpin2018,lebrunLongdistanceSpintransportMorin2020,hanBirefringencelikeSpinTransport2020,hoogeboomNonlocalSpinSeebeck2020}. Compared with easy-axis AFMs, such easy-plane AFMs exhibit highly tunable spin dynamics \cite{kalashnikovaImpulsiveGenerationCoherent2007,satohSpinOscillationsAntiferromagnetic2010,tzschaschelUltrafastOpticalExcitation2017} and potentially a form of dissipationless spin transport known as spin superfluidity \cite{soninSpinCurrentsSpin2010,soninSuperfluidSpinTransport2020}. 

A recent study of spin propagation performed on the easy-plane  antiferromagnet CrSBr \cite{baeExcitoncoupledCoherentMagnons2022} vividly illustrates the potential of this class of materials. Bae et al. \cite{baeExcitoncoupledCoherentMagnons2022} demonstrated that the dynamics of spin waves, whose characteristic energies are below 1 meV, can be probed by $\approx$ 1 eV photons as a consequence of strong exciton-magnon interaction -- a result with implications for the transduction of quantum information. However, a striking puzzle emerged when the group velocity, $v_g$, of spin wavepackets was compared with that expected from inelastic neutron scattering (INS) \cite{scheieSpinWavesMagnetic2022}; $v_g$ was found to be orders of magnitude larger than the velocity estimated by extrapolating the magnon dispersion measured by INS to the small wavevectors that comprise optically generated spin wavepackets.  As a possible explanation, it was suggested that long-wavelength magnons in CrSBr can propagate close to the sound velocity through coupling to acoustic phonons.  
\begin{figure}[tb]
\centering
\includegraphics[width=0.52\columnwidth]{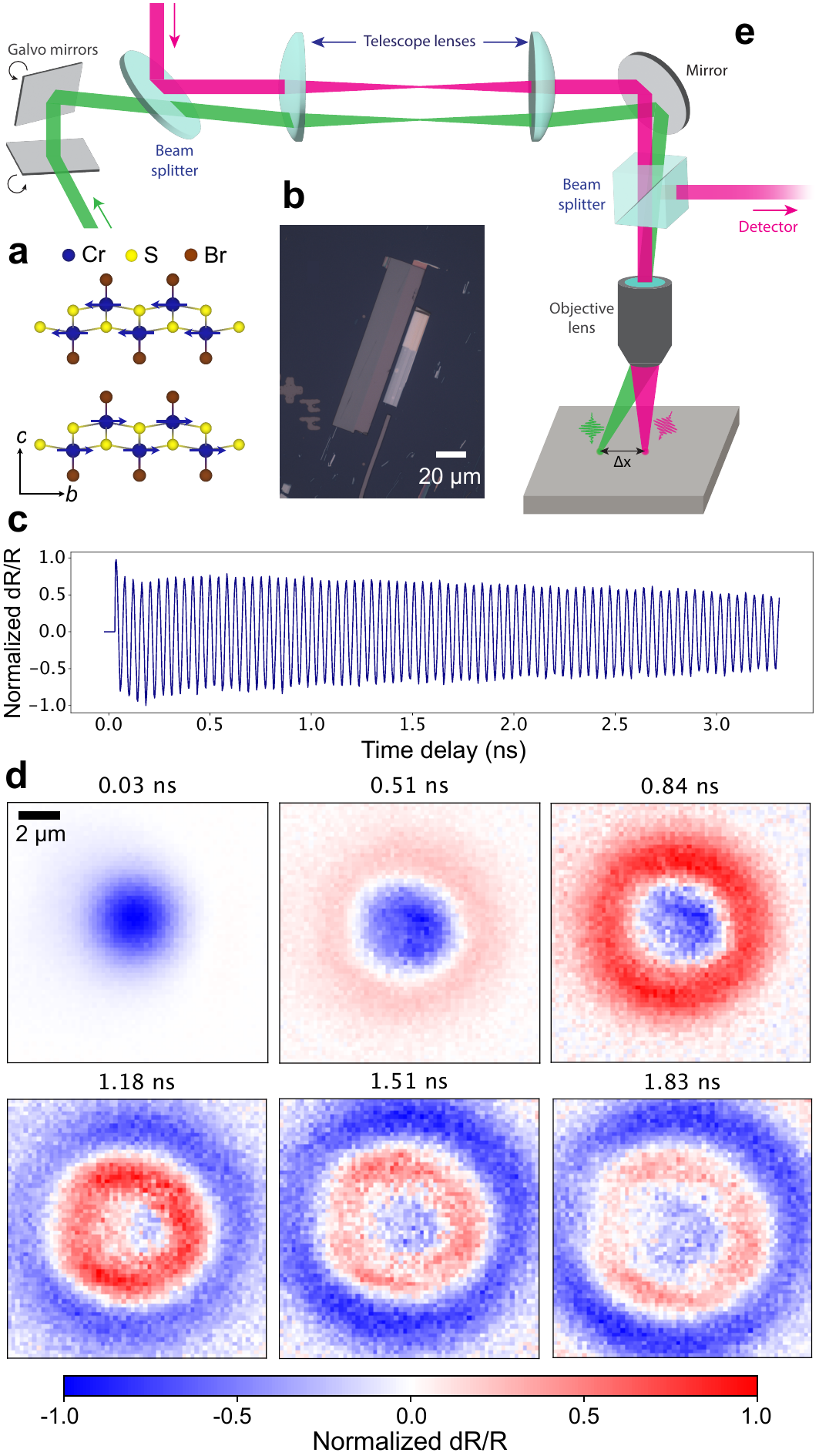}
\caption{(a) Crystal structure of CrSBr viewed along the a-axis. The blue arrows indicate the orientation of Cr spins; (b) Optical image of exfoliated CrSBr flakes; (c) Transient reflectance measured on a 20 nm thick crystal with overlapping pump and probe beams; (d) Snapshots of spatially resolved transient reflectance measured on a 332 nm thick crystal. The images were obtained by rastering the pump beam and sampling the reflectance at fixed time delays between pump and probe pulses. The measurements shown in panels (c) and (d) were performed at a temperature 2.5 K and a field of 1.0 T applied along the $c$ direction, using pump and probe wavelengths of 700 and 910 nm, respectively. The diameters of the pump and probe focus were $\approx$ 2.0 $\mu$m). (e) Schematic of the experimental setup. Two orthogonal galvo-driven mirrors and a 4f telescope enable automated scanning of the pump laser beam.}
\label{fig:Fig1_mapping}
\end{figure}

Here, we show that the resolution of this puzzle is that spin wavepacket propagation in CrSBr arises not from magnetoelastic coupling, but from the long-range magnetic dipole interaction, which is often ignored in the context of AFMs. This conclusion is based on the discovery of several salient features of the spin dynamics: (1) the near isotropy of spin wave propagation; (2) the dependence of $v_{g}$ on sample thickness and applied magnetic field; (3) the observation that the group and phase velocity of the spin wavepacket are opposite in sign; and (4) the magnon density of states.  Each of these features is quantitatively accounted for by a theory of dipolar spin waves whose parameters are obtained from independent equilibrium measurements.  Finally, we demonstrate that our theoretical understanding of the propagation mechanism enables optimization of applied field, sample thickness, and focal spot size to demonstrate spin wavepacket propagation over tens of microns. 

\section{Detection of coherent magnon propagation}
CrSBr is a 2D van der Waals semiconductor with easy-plane layered antiferromagnetic order, as depicted in Fig.~\ref{fig:Fig1_mapping}a. In addition to the easy-plane anisotropy, the spins experience a weaker anisotropy within the plane, with the easy axis in the $b$ direction. This anisotropy is consistent with the orthorhombic crystal structure, which is reflected in the rectangular faceting of the exfoliated thin flakes shown in Fig.~\ref{fig:Fig1_mapping}b. Spin waves in CrSBr are photoexcited when absorption of a light pulse generates a transient change in this equilibrium spin configuration. As demonstrated in Ref.~\cite{baeExcitoncoupledCoherentMagnons2022} the spin wave amplitude is enhanced by the application of a static magnetic field, which tips the spins from their in-plane, easy-axis orientation. The ensuing spin wave oscillations efficiently modulate the exciton energy, enabling magnon detection via measurement of transient reflectance at photon energies near the exciton absorption peak 1.36 eV.  

Fig.~\ref{fig:Fig1_mapping}c shows the highly underdamped oscillations in transient reflectance seen when the position of the pump and probe beams coincide. In this example, the measurement was performed on a 20 nm thick crystal at a temperature 2.5 K and a field of 1.0 T applied along the $c$ direction. Although two magnon modes are expected for a biaxial AFM such as CrSBr, only a single frequency of 21 GHz is observed. This is consistent with a previous report \cite{diederichTunableInteractionExcitons2023} showing that for magnetic fields applied along the symmetry directions $H\parallel c$ or $a$ only one magnon mode makes a first-order contribution to the spin-dependent exciton energy. For $H\parallel c$, the bright mode is the low-frequency magnon branch.   

Fig. \ref{fig:Fig1_mapping}d shows snapshots of the spin wave amplitude at various time delays ranging from 0.03 to 1.83 ns after pulsed photoexcitation. The maps were obtained using a transient reflectance microscope with spatial scanning capability, shown schematically in Fig.~\ref{fig:Fig1_mapping}e. The images show that the spin wavepacket propagates with nearly isotropic group velocity, reaching distances of $\approx$ 5 $\mu$m after $\approx$ 2 ns, and provide unambiguous evidence of coherent magnon propagation. These results are essentially different from the highly anisotropic, quadrupolar pattern reported previously \cite{baeExcitoncoupledCoherentMagnons2022} and cited as a evidence for propagation via magnetoelastic coupling. Given that two magnon frequencies were observed in Ref. \cite{baeExcitoncoupledCoherentMagnons2022} it is likely that the quadrupolar pattern results from the superposition of two magnon branches with different anisotropies and group velocities. 

\begin{figure*}[t]
\centering
\includegraphics[width=1.0\columnwidth]{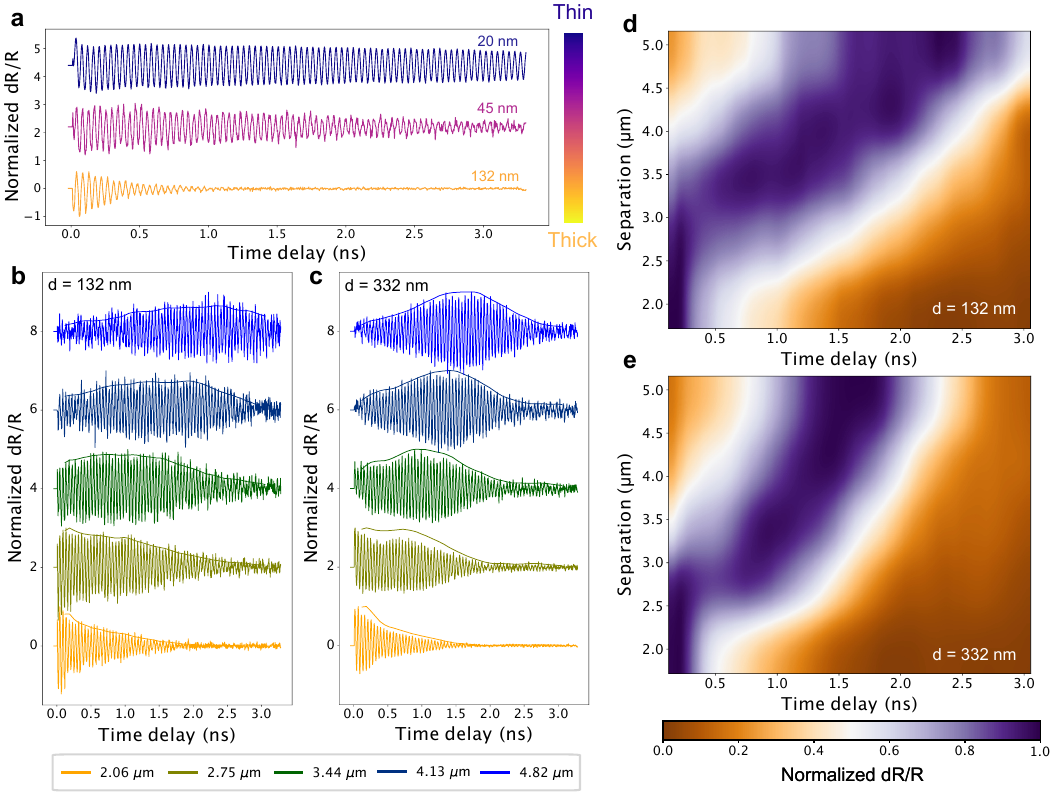}
\caption{(a) Transient reflectance with pump and probe beams overlapped for crystals with thicknesses of 20, 45 and 132 nm; (b,c) Transient relfectance measured for pump-probe separations ranging from 2.06 to 4.82 $\mu$m in samples with thickness (b) 132 nm and (c) 332 nm. Solid lines show the envelope function of each time trace; (d,e) The envelope amplitude plotted in the plane of time-delay (vertical axis) and pump-probe separation (horizontal axis).}
\label{fig:Fig2_thickness}
\end{figure*}

The first, qualitative observation that points directly to propagation via the magnetic dipole interaction is the pronounced dependence of the dynamics on sample thickness, $d$, over a broad range from 20-400 nm. Fig.~\ref{fig:Fig2_thickness}a shows the transient reflectance with pump and probe beams overlapped for crystals with $d$ = 20, 45 and 132 nm. The decay time is seen to   decrease dramatically with increasing thickness. Although one might at first infer that the spin wave damping increases with increasing $d$, this is, in fact, not the case. As we show below, when pump and probe beams coincide the decrease of the amplitude with time reflects the $d$-dependent velocity at which the spin wavepacket propagates away its location at time $t=0$. While a dependence of $v_g$ on $d$ in this regime is difficult to reconcile with a local interaction such as exchange or spin-phonon coupling, it arises naturally from the long-range nature of dipole coupling.  

To quantify the dependence of group velocity on crystal thickness, we characterized the propagation in samples with $d$ = 132 and 332 nm. We determine $v_g$ for each of the samples from transient reflectance measured for five values of pump-probe separation along the a-axis (Figs.~\ref{fig:Fig2_thickness}(b,c)). The solid lines show the envelope function of the propagating wavepackets. It is clear from the time-dependence at different positions that the amplitude measured for coincident pump and probe decreases because of propagation away from the point of origin, rather than from dissipation. The contrast in propagation dynamics for the two samples is illustrated in Figs.~\ref{fig:Fig2_thickness}(d,e), in which the amplitude of the spin wavepacket envelope is plotted with vertical and horizontal axes corresponding to pump-probe separation and time delay, respectively. The slope of the purple stripe captures $v_{g}$, while its width is a measure of spin wavepacket spreading. From the slope, we infer group velocities of 1.1$\pm$0.1 km/s and 2.5$\pm$0.1 km/s for the 132 nm and 332 nm crystals, respectively; the ratio of group velocities matches the ratio of thicknesses. A linear dependence of $v_g$ on $d$ will emerge as a feature of the theoretical analysis of spin wave dispersion in the long wavelength regime described below. 

\section{Spin waves in C\MakeLowercase{r}SB\MakeLowercase{r}}
\subsection{Theory}
In ferromagnets, spin waves in the long-wavelength dipole-mediated regime have been well-recognized and studied \cite{damonMagnetostaticModesFerromagnet1961, hurbenTheoryMagnetostaticWaves1996, satohDirectionalControlSpinwave2012, demokritovBoseEinsteinCondensation2006,matsumotoTheoreticalPredictionRotating2011}, and are known as magnetostatic waves (MSWs).  Although there is no net magnetization in equilibrium, long-range dipole coupling arises in AFMs as well from dynamic fluctuations out of equilibrium. In the following we extend the theory of dipolar modes in AFMs \cite{camleyLongWavelengthSurfaceSpin1980,luthiSurfaceSpinWaves1983,stampsMagnetostaticModesThin1986} to van der Waals magnets such as CrSBr, where the existence of both out-of-plane and in-plane anisotropy plays a critical role. We consider alternating layers of magnetization, $\bm{M}_{1,2}$, that are antiparallel in equilibrium . The Landau-Lifshitz equations for the dynamics of the magnetic sublattices,  
\begin{equation}
\label{LL}
    \frac{\partial \bm{M}_i}{\partial t}=- \gamma \bm{M}_i\times \bm{H}^{\textrm{eff}}_i,
\end{equation}
combined with Maxwell's equations in the magnetostatic regime, $\nabla \cdot \bm{B} =\nabla \times \bm{H} = 0$, form a closed set that yield the spin collective mode frequencies and eigenvectors. In Eq.~\ref{LL}, $\gamma$ is the gyromagnetic ratio and $\bm{H}^{\textrm{eff}}_i$ is the sum of effective fields arising from anisotropy, interlayer exchange and the dynamical magnetic field $\bm{h}(t)$. The effective field on each sublattice is given by,
\begin{equation}
  \bm{H}^{\textrm{eff}}_i=-\frac{\partial F}{\partial \bm{M}_i}+\bm{h}_i(t), 
\end{equation}
where $F$ is the sum of exchange and anisotropy contributions, i.e. $F=F_{ex}+F_a$. The exchange term is,
\begin{equation}
\label{eq:exchange}
F_{ex}=J\frac{\bm{M}_{1}\cdot\bm{M}_{2}}{M_{s}^{2}},
\end{equation}
where $J$ is the exchange constant ($J>0$) and $M_{s}$ is the saturation magnetization of each sublattice. The easy-plane anisotropy is expressed as,
\begin{equation}
\label{eq:anisotropy}
F_{a} = K_{z}\frac{M_{1z}^{2}+M_{2z}^{2}}{M_{s}^{2}}-K_{x}\frac{M_{1x}^{2}+M_{2x}^2}{M_s^2},
\end{equation}
where $K_{x}$ and $K_{z}$ are anisotropy constants ($K_{x}, K_{z}>0$). The first term on the right-hand side of Eq. \ref{eq:anisotropy} confines the spins to the $xy$ plane while the second term expresses the easy-axis anisotropy within the plane. Although $K_x$ is typically much smaller than $K_z$, it plays an important role in the dispersion of long-wavelength spin waves.

Within an $x$-oriented domain the equilibrium magnetization is $\bm{M}_{1}=(M_{s},0,0)$ and $\bm{M}_{2}=(-M_{s},0,0)$ and small fluctuations from equilibrium are transverse, i.e. $\bm{m}_i=(m_{iy},m_{iz})$. Assuming solutions of the form $\bm{m}_i(\bm{r},t)=\bm{m_{i0}}e^{i(\bm{k}\cdot\bm{r}-\omega t)}$, we find two spin wave bands with dispersion given by,
\begin{widetext}
\begin{equation}
\label{eq:dispersion}
\omega_{\pm}(\bm{\hat{k}})=\sqrt{
    \omega_{T}\omega_{x}+\frac{\omega_{J}(\omega_{T}+\omega_{x})}{2}+\frac{\omega_{M}(k_{y}^{2}\omega_{T}+k_{z}^{2}\omega_{x})}{k^{2}}
    \pm \frac{1}{2}A(\hat{\bm{k}})
    },
\end{equation}
\begin{equation}
\label{A(k)}
    A(\hat{\bm{k}})= \sqrt{\omega_{J}^{2}(\omega_{T}-\omega_{x})^{2}+4\omega_{J}\omega_{M}(\omega_{T}-\omega_{x})\left(\frac{\omega_{T}k_{y}^{2}-\omega_{x}k_{z}^{2}}{k^2}\right)+4\omega_{M}^{2}\left(\frac{\omega_{T}k_{y}^{2}+\omega_{x}k_{z}^{2}}{k^2}\right)^{2}},
\end{equation}
\end{widetext}
where $\omega_{M}=4\pi\gamma M_{s}$, $\omega_{x}=2\gamma K_x/M_s$, $\omega_{T}=2\gamma(K_{x}+K_{z})/{M_{s}}$,  $\hat{\bm{k}}=\bm{k}/k$ (see Supplementary Information for derivation of the dispersion relations). The spin wave frequencies depend only on the direction of the wavevector $\bm{k}$ and not its magnitude -- a property that arises from the long wavelength nature of the interaction \cite{camleyLongWavelengthSurfaceSpin1980}. Because the dispersion is a function of $k_{x}/k_{z}$ and $k_{y}/k_{z}$, the group velocity in the $xy$ plane is proportional to $1/k_{z}$ and therefore to $d$, which has been proved in the thickness dependent measurements (Fig.~\ref{fig:Fig2_thickness}) above.

To consider propagation in the plane of thin films or flakes, we focus on the dispersion of $\omega_{\pm}$ in the $k_x,k_y$ plane, which is shown in Fig.~\ref{fig:Fig3_theory}a. Fig.~\ref{fig:Fig3_theory}b shows the dispersion along the $k_x=0$ and $k_y=0$ directions, illustrating that propagation is predicted to have qualitatively different properties in the two magnon branches. The lower energy band disperses in all directions in the plane and is ``backwards moving'' in the sense that its phase and group velocity are opposite in sign. By contrast, the higher energy band is strongly dispersive and forward-moving for propagation in the $y$ direction (perpendicular to the equilibrium N\'{e}el vector), but exhibits no dispersion in the direction parallel to the equilibrium N\'{e}el vector. 

\subsection{Testing the theoretical predictions}
\subsubsection{Negative group velocity}

The prediction of negative group velocity of the low-frequency spin wave branch, $\omega_{-}$ was tested by measuring the transient reflectance as a function of pump-probe separation along both $a$- and $b$-axes. Fig.~\ref{fig:Fig3_theory}c shows results for the $a$-axis, with amplitude represented by color in the time separation plane (the data for $b$-axis propagation are very nearly the same, see Supplementary Information). The negative slope of the lines of constant phase confirms that the $\omega_{-}$ mode is backward propagating, in agreement with our theoretical prediction for the easy-plane, biaxial AFM. The observation that $a$ and $b$ directions have the same sign of group velocity is consistent with the nearly isotropic rings seen in the maps of Fig.~\ref{fig:Fig1_mapping}e. We suggest that quadrupolar pattern, reported in Ref.~\cite{baeExcitoncoupledCoherentMagnons2022}, results from simultaneous excitation of the lower ($v_g<0$) and upper ($v_g>0$) branches, which occurs when the applied magnetic field has both in and out-of-plane components.

\subsubsection{Field dependence of $v_{g}$}
\begin{figure}[t]
\centering
\includegraphics[width=0.8\columnwidth]{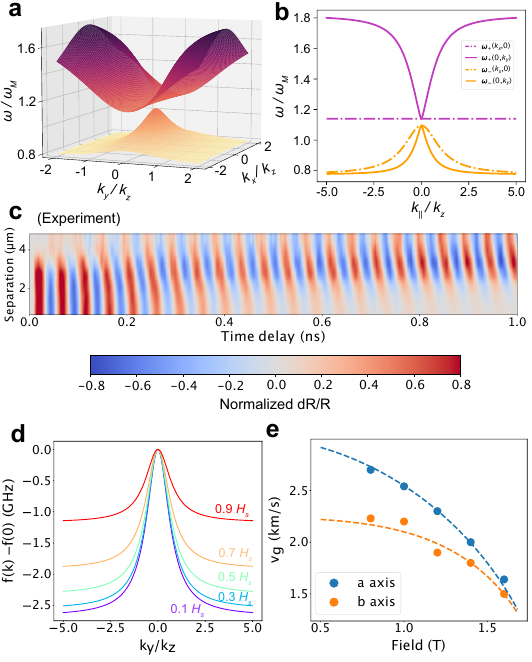}
\caption{(a) Dispersion of the two magnon branches in the $k_x,k_y$ plane; (b) Dispersion along the $k_{x}$ axis (dash-dotted lines) and the $k_{y}$ axis (solid lines) for both $\omega_{-}$ (orange lines) and $\omega_{+}$ (purple lines) branches calculated from Eq. \eqref{eq:dispersion} and parameters $\omega_{T}/\omega_{M} = 1.0$, $\omega_{x}/\omega_{M} = 0.3$ and $\omega_{J}/\omega_{M} = 1.0$; (c) Plot of the transient reflectance in the time-separation plane measured along the $a$-axis; (d) Linecuts of the $\omega_{-}$ mode dispersion along the $k_{y}$ direction at external fields ranging from $0.1 H_{s}$ to $0.9 H_{s}$; (e) Group velocity $v_{g}$ along both $a$- and $b$-axes for the $\omega_{-}$ mode as a function of external field. The solid dots are the values of $v_{g}$ extracted from the experimental data and the dash lines are the theoretical predictions.}
\label{fig:Fig3_theory}
\end{figure}

A point worth emphasizing is that our MSW theory for spin wave dispersion in a biaxial AFM includes no free parameters; all quantitative predictions follow from terms in the free energy ($K_{x} = 2.00\times 10^{4}$ J/m$^{3}$, $K_{z} = 1.10\times 10^{5}$ J/m$^{3}$, $J = 6.48\times 10^{4}$ J/m$^{3}$ and $M_{s}=2.05\times 10^{5}$ A/m) that are determined independently from the magnetic field dependence of the spin wave frequency. An example of such a quantitative prediction is the dependence of $v_g$ on magnetic field applied along the $c$-axis. As shown in Fig.~\ref{fig:Fig3_theory}d the dispersion of the $\omega_{-}$ mode is predicted to gradually flatten with increasing applied field, with $v_{g}$ expected to decrease by more than 50\% as the applied field approaching $H_{s}$. Fig.~\ref{fig:Fig3_theory}e shows $v_g$ measured as a function of $H$ for both $a$- and $b$-axes, using the method introduced in Fig.~\ref{fig:Fig2_thickness}. The solid lines in the figure are the no-free parameter predictions of the theory, which accurately correspond to the experimental values, plotted as solid dots.

\subsubsection{Magnon density states}
\begin{figure}[t]
\centering
\includegraphics[width=1.0\columnwidth]{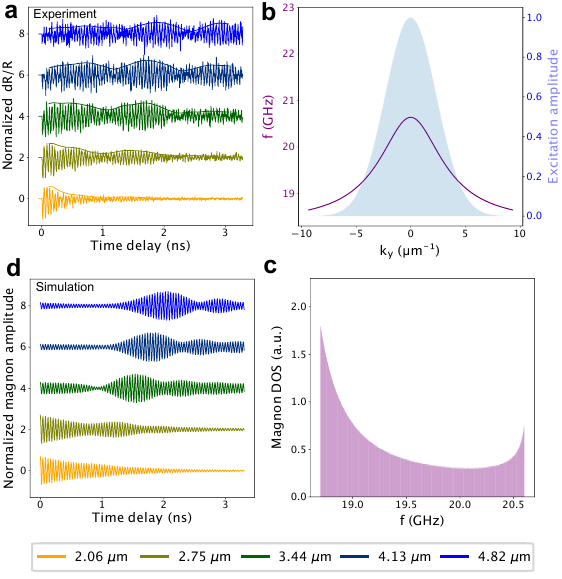}
\caption{(a) Wavepacket propagation under the same conditions as Fig. \ref{fig:Fig2_thickness}c, with diameter of the pump beam focus reduced to 2 to 1.0 $\mu$m. Beating pattern emerges in the amplitude of the wavepacket envelope; (b) The shaded region shows the distribution of wavevectors that are excited by the 1 $\mu$m diameter laser focus on the same scale as the predicted magnon dispersion; (c) Magnon density of states corresponding to the dispersion in panel (b); (d) Theoretically predicted magnon wavepackets.}
\label{fig:Fig4_DOS}
\end{figure}

Another important prediction of the MSW theory, evident in Fig.~\ref{fig:Fig3_theory}d, is that the magnon dispersion relation is strongly nonlinear, even on the scale of wavevectors probed optically.  Anomalous spin wavepacket propagation attributable to nonlinear MSW dispersion has been reported recently in a ferromagnet in which spin dynamics are damped \cite{leeSpinWavepacketsKagome2023}. The influence of nonlinear dispersion on propagation in CrSBr, where spin waves are very weakly damped, is even more striking. Fig.~\ref{fig:Fig4_DOS}a illustrates wavepacket propagation under the same conditions as Fig. \ref{fig:Fig2_thickness}c, except that the diameter of the pump beam focus has been reduced from 2 $\mu$m to 1.0 $\mu$m. Remarkably, with the tighter focus, clear beating patterns emerge in the time traces measured at several pump-probe separations. 

The qualitative origin of the beating pattern is illustrated in Figs.~\ref{fig:Fig4_DOS}b and c. The shaded region in Fig.~\ref{fig:Fig4_DOS}b shows the distribution of wavevectors that are excited by the 1 $\mu$m diameter laser focus and the solid line shows the predicted magnon dispersion relation in the same wavevector range. Compared to the 2 $\mu$m diameter pump beam focus, the 1 $\mu$m diameter one covers a wider range of $k$-space. The propagation of the wavepacket is governed by its spectral content, which in turn which reflects the magnon density of states (DOS). The strongly nonlinear dispersion relation maps to the bimodal DOS shown in Fig.~\ref{fig:Fig4_DOS}c, whose two peaks give rise to the observed beating pattern of the spin wave oscillation. For a quantitative comparison of theory and experiment, we calculated the time evolution of magnon wavepackets using the relation,
\begin{equation}
\label{eq:Integration}
\bm{m}_i(x,y,t)=\bm{m_{i}}\int d^{2}\bm{k}\ e^{-k^{2}\sigma^{2}/2} e^{-i\omega_{-}(\bm{k})t} e^{i\bm{k}\cdot \bm{r}},
\end{equation}
where $\bm{m}_i(x,y,t)$ is the departure of the magnetization from equilibrium, $\bm{k}=(k_{x},k_{y})$, $\bm{r}=(x,y)$ and $\sigma$ is the laser spot size. The spin wavepackets computed using Eq. \eqref{eq:Integration}, for a 1.0 $\mu$m focus and the same anisotropy parameters introduced previously, are shown Fig.~\ref{fig:Fig4_DOS}d. The oscillatory envelope function determined from the theoretical MSW dispersion is in excellent agreement with the experimental results.

\begin{figure}[t]
\centering
\includegraphics[width=1.0\columnwidth]{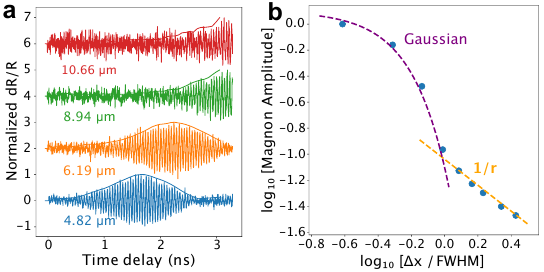}
\caption{(a) Transient reflectance measured for pump-probe separation ranging from 4.82 to 10.66 $\mu$m in a sample with 372 nm thickness; (b) A double logarithmic plot of the amplitude of the wavepacket
vs. pump-probe separation. For separations where pump and probe beams overlap the amplitude decreases as a Gaussian. For nonoverlapping separations the amplitude falls off in proportion to $1/r$.}
\label{fig:Fig5_long_distance}
\end{figure}

\subsubsection{The range of coherent magnon propagation}
A surprising consequence of the long-range nature of dipole coupling is that spin wavepacket propagation is sensitive to the macroscopic geometry of the crystal.  As a consequence, coherent spin propagation is optimized when two characteristic length scales: sample thickness and diameter of the pump focus, are approximately equal. To see why, consider the inverted V-shape dispersion of the lower branch (Fig. \ref{fig:Fig3_theory}d), which exhibits a thickness-dependent crossover from linear dispersion to a nearly flat band at an in-plane wavevector, $k_c\approx 1/d$.  On the other hand the total bandwidth of MSWs is independent of $d$ and is given by $\delta\equiv\sqrt{(2\omega_M+\omega_J+\omega_T)\omega_x}-\sqrt{(\omega_J+\omega_T)\omega_x}$, from which it follows that $v_g\approx d\delta$ for $k<k_c$ and $v_g\approx 0$ for $k>k_c$.  The diameter of the pump focus, $\sigma$, determines the range of wavevectors, $\Delta k$, that comprise the wavepacket.  In the large thickness limit, $\Delta k d\gg1$, the group velocity is large, but only a small fraction of the excited wavevectors fall in the propagating regime of the dispersion relation.  In the opposite limit, $\Delta kd \ll 1$, although the entire range of excited wavevectors are propagating, the group velocity is small.  

To find the ultimate range of coherent spin propagation in CrSBr, we performed measurements in the optimal regime, $\Delta k d \approx 1$, obtained by choosing $\sigma=2 \mu$m and $d=372$ nm. As shown in Fig.~\ref{fig:Fig5_long_distance}a, oscillations in transient reflectance are clearly observed at a pump-probe separation, $\Delta x$, of 10.66 $\mu$m. Although this distance is already impressively long, it does not represent an upper bound, as the wavepacket clearly continues to propagate beyond our time-measurement window of 3 ns. In another attempt to determine an upper bound, we evaluated the integral of the envelope function for $0\leqslant \Delta x\leqslant 6\ \mu$m, the window in which we can resolve the entire wavepacket. In Fig.~\ref{fig:Fig5_long_distance}b, we plot the integrated amplitude as a function of $\Delta x$ in a log-log scale. For $\Delta x\leqslant \sigma$ the amplitude decays as $\exp[-(\Delta x/\sigma)^2]$ reflecting the spatial overlap of the pump and probe foci. For larger separations the magnon amplitude varies as $1/r$, which means that we do not detect any deviation from coherent, non-dissipative propagation to a distance $\approx 6\mu$m. Finally, although we cannot determine an upper bound on the coherent propagation distance, we can establish a lower bound by considering the dynamics in our thinnest sample. From the data in Fig. \ref{fig:Fig1_mapping}c, we infer a minimum decay time of 5.3 ns, if we attribute the entire decrease in amplitude to dissipation, rather than propagation. This leads to a lower bound on the characteristic length scale for coherent propagation $l\approx 3$ km/s$\times$ 5 ns $=$ 15 $\mu$m and the possibility of detecting propagation to much larger distances considering the high sensitivity of optical readout via magnon-exciton coupling.

\section{Conclusion}

In summary, we have demonstrated that the mechanism for optically injected spin wavepacket propagation in CrSBr is long-range coupling among Cr spins, arising from their magnetic-dipole interaction rather than short-range exchange. This conclusion is validated by quantitative agreement between theoretical predictions and experimental findings for the dependence of group velocity on propagation direction, crystal thickness and external magnetic field, as well as its (negative) sign. In as much as spin dynamics in this regime can be quantitatively predicted based only on the Landau-Lifshitz equation and Maxwell’s equations in the quasistatic limit, we anticipate that propagation via magnetostatic modes is a universal feature of the vdW AFMs, currently under intense study. 

Our findings have implications beyond identifying the mechanism for spin transport, as they indicate areas for caution and for promise in future investigations of the vdW magnets. The reason for caution is that in the magnetostatic regime coherent propagation cannot be realized in few-layer crystals, as $v_g\rightarrow 0$ in this limit. The promise is that propagation can be optimized by tuning the range of wavevectors in an optically or electrically injected wavepacket to be on the order of the inverse sample thickness.  With such optimization and within our time window of 3 ns, we could not detect any departure from coherent, ballistic magnon propagation, although a lower bound for $r^{-1}$ decrease in wavepacket amplitude of 15 $\mu$m was found.  As a final comment, enabling spin propagation in the few layer limit requires tuning anisotropy towards zero, where local-exchange interactions will begin to dominate the dynamics. With such tuning magnetic systems enter a regime of strong thermal and quantum fluctuations in which the nature of spin propagation is an exciting area for future research.

\section{Methods}
\subsection{Crystal synthesis and sample preparation}
CrSBr crystals were prepared by a direct reaction from for elements. Chromium (99.99\%, -60 mesh, Chemsavers, USA), bromine (99.9999\%, Sigma-Aldrich, Czech Republic), and sulfur (99.9999\%, Stanford Materials, USA) were mixed in a stochiometric ratio in a quartz ampoule (35 $\times$ 220 mm) corresponding to 15 g of CrSBr. Bromine excess of 0.5 g was used to enhance vapor transport. The material was pre-reacted in an ampoule using a crucible furnace at 700 $^{\circ}$C for 12 h, while the second end of the ampoule was kept below 250 $^{\circ}$C. The heating procedure was repeated two times until the liquid bromine disappeared. The ampoule was placed in a horizontal two-zone furnace for crystal growth. First, the growth zone was heated to 900 $^{\circ}$C, while the source zone was heated at 700 $^{\circ}$C for 25 h. For the growth, the thermal gradient was reversed and the source zone was heated from 900 to 940 $^{\circ}$C and the growth zone from 850 to 800 $^{\circ}$C over a period of 7 days. The crystals with dimensions up to 5 $\times$ 20 mm were removed from the ampule in an Ar glovebox.

Following synthesis, bulk crystals were mechanically exfoliated onto a silicon wafer with a 90-nm-thick silicon dioxide (SiO$_2$) layer inside a glove box. Atomic force microscope images of the CrSBr flakes were taken to identify the sample thicknesses.

\subsection{Transient reflectance microscope}
The transient reflectance experiments were carried out with 700-nm pump and 910-nm probe pulses generated from the ORPHEUS-TWINS optical parametric amplifiers pumped by the Light Conversion CARBIDE Yb-KGW laser amplifier operating at the repetition rate of 300 kHz. Both beams were focused onto the sample surface with approximate spot sizes of 2 $\mu$m unless otherwise stated, with incident laser powers fixed at $\approx$ 1 $\mu$W. The position of the pump focus was scanned by adjusting the voltage applied to the 2-axis galvanometer-driven mirrors, which are located at a distance 4f (f = 50 cm) before the entrance aperture of the final 50x objective lens (N.A. = 0.50). A pair of telescope lenses with focal lengths of f are placed equidistant from the galvo mirrors and the objective so that the laser beam steered from the galvo mirrors forms a one-to-one image at the entrance of the objective lens. The pump laser pulses were modulated at 10 kHz with a chopper and the transient reflectance signals were measured with a lock-in amplifier (MFLI, Zurich Instruments). The external magnetic field was applied by the superconducting coil inside the Quantum Design OptiCool cryostat.

\section{Acknowledgements} 
We acknowledge support of the Quantum Materials program under the Director, Office of Science, Office of Basic Energy Sciences, Materials Sciences and Engineering Division, of the U.S. Department of Energy, Contract No. DE-AC02-05CH11231. J.O and Y.S received support from the Gordon and Betty Moore Foundation's EPiQS Initiative through Grant GBMF4537 to J.O. at UC Berkeley. F.M. and J.Y. acknowledge support from the U.S. Department of Energy, Office of Science, Office of Basic Energy Sciences, Materials Sciences and Engineering Division under contract DE-AC02-05-CH11231 (Organic-Inorganic Nanocomposites KC3104). Z.S. was supported by ERC-CZ program (project LL2101) from Ministry of Education Youth and Sports (MEYS).

\section{Author contributions}
Y.S. and J.O. designed research. Y.S. carried out all optical measurements with assistance from C.L. under the supervision of J.O. Bulk crystals were synthesized and characterized by A.S. under the supervision of Z.S. F.M. prepared and characterized thin flakes under the supervision of J.Y. H.Z. and F.M. performed atomic force microscope measurements under the supervision of R.R. Theoretical analysis was performed by Y.S. and J.O. Y.S. and J.O. wrote the paper.

\section{Competing interests}
The authors declare no competing interests.

\section{Data Availability}
All data sets supporting the conclusions of the paper and Supplementary Information are shared in a public accessible repository.

\section{Code Availability}
The computer codes used to generate results are provided in the Supplementary Information.

\bibstyle{apsrev4-1}

\bibliography{Main_text_citation}

\begin{thebibliography}{31}%
\makeatletter
\providecommand \@ifxundefined [1]{%
 \@ifx{#1\undefined}
}%
\providecommand \@ifnum [1]{%
 \ifnum #1\expandafter \@firstoftwo
 \else \expandafter \@secondoftwo
 \fi
}%
\providecommand \@ifx [1]{%
 \ifx #1\expandafter \@firstoftwo
 \else \expandafter \@secondoftwo
 \fi
}%
\providecommand \natexlab [1]{#1}%
\providecommand \enquote  [1]{``#1''}%
\providecommand \bibnamefont  [1]{#1}%
\providecommand \bibfnamefont [1]{#1}%
\providecommand \citenamefont [1]{#1}%
\providecommand \href@noop [0]{\@secondoftwo}%
\providecommand \href [0]{\begingroup \@sanitize@url \@href}%
\providecommand \@href[1]{\@@startlink{#1}\@@href}%
\providecommand \@@href[1]{\endgroup#1\@@endlink}%
\providecommand \@sanitize@url [0]{\catcode `\\12\catcode `\$12\catcode
  `\&12\catcode `\#12\catcode `\^12\catcode `\_12\catcode `\%12\relax}%
\providecommand \@@startlink[1]{}%
\providecommand \@@endlink[0]{}%
\providecommand \url  [0]{\begingroup\@sanitize@url \@url }%
\providecommand \@url [1]{\endgroup\@href {#1}{\urlprefix }}%
\providecommand \urlprefix  [0]{URL }%
\providecommand \Eprint [0]{\href }%
\providecommand \doibase [0]{https://doi.org/}%
\providecommand \selectlanguage [0]{\@gobble}%
\providecommand \bibinfo  [0]{\@secondoftwo}%
\providecommand \bibfield  [0]{\@secondoftwo}%
\providecommand \translation [1]{[#1]}%
\providecommand \BibitemOpen [0]{}%
\providecommand \bibitemStop [0]{}%
\providecommand \bibitemNoStop [0]{.\EOS\space}%
\providecommand \EOS [0]{\spacefactor3000\relax}%
\providecommand \BibitemShut  [1]{\csname bibitem#1\endcsname}%
\let\auto@bib@innerbib\@empty
%</preamble>
\bibitem [{\citenamefont {Gibertini}\ \emph {et~al.}(2019)\citenamefont
  {Gibertini}, \citenamefont {Koperski}, \citenamefont {Morpurgo},\ and\
  \citenamefont {Novoselov}}]{gibertiniMagnetic2DMaterials2019a}%
  \BibitemOpen
  \bibfield  {author} {\bibinfo {author} {\bibfnamefont {M.}~\bibnamefont
  {Gibertini}}, \bibinfo {author} {\bibfnamefont {M.}~\bibnamefont {Koperski}},
  \bibinfo {author} {\bibfnamefont {A.~F.}\ \bibnamefont {Morpurgo}},\ and\
  \bibinfo {author} {\bibfnamefont {K.~S.}\ \bibnamefont {Novoselov}},\ }\href
  {https://doi.org/10.1038/s41565-019-0438-6} {\bibfield  {journal} {\bibinfo
  {journal} {Nature Nanotechnology}\ }\textbf {\bibinfo {volume} {14}},\
  \bibinfo {pages} {408} (\bibinfo {year} {2019})}\BibitemShut {NoStop}%
\bibitem [{\citenamefont {Gong}\ and\ \citenamefont
  {Zhang}(2019)}]{gongTwodimensionalMagneticCrystals2019a}%
  \BibitemOpen
  \bibfield  {author} {\bibinfo {author} {\bibfnamefont {C.}~\bibnamefont
  {Gong}}\ and\ \bibinfo {author} {\bibfnamefont {X.}~\bibnamefont {Zhang}},\
  }\href {https://doi.org/10.1126/science.aav4450} {\bibfield  {journal}
  {\bibinfo  {journal} {Science}\ }\textbf {\bibinfo {volume} {363}},\ \bibinfo
  {pages} {eaav4450} (\bibinfo {year} {2019})}\BibitemShut {NoStop}%
\bibitem [{\citenamefont {Mak}\ \emph {et~al.}(2019)\citenamefont {Mak},
  \citenamefont {Shan},\ and\ \citenamefont
  {Ralph}}]{makProbingControllingMagnetic2019a}%
  \BibitemOpen
  \bibfield  {author} {\bibinfo {author} {\bibfnamefont {K.~F.}\ \bibnamefont
  {Mak}}, \bibinfo {author} {\bibfnamefont {J.}~\bibnamefont {Shan}},\ and\
  \bibinfo {author} {\bibfnamefont {D.~C.}\ \bibnamefont {Ralph}},\ }\href
  {https://doi.org/10.1038/s42254-019-0110-y} {\bibfield  {journal} {\bibinfo
  {journal} {Nature Reviews Physics}\ }\textbf {\bibinfo {volume} {1}},\
  \bibinfo {pages} {646} (\bibinfo {year} {2019})}\BibitemShut {NoStop}%
\bibitem [{\citenamefont {Bae}\ \emph {et~al.}(2022)\citenamefont {Bae},
  \citenamefont {Wang}, \citenamefont {Scheie}, \citenamefont {Xu},
  \citenamefont {Chica}, \citenamefont {Diederich}, \citenamefont {Cenker},
  \citenamefont {Ziebel}, \citenamefont {Bai}, \citenamefont {Ren},
  \citenamefont {Dean}, \citenamefont {Delor}, \citenamefont {Xu},
  \citenamefont {Roy}, \citenamefont {Kent},\ and\ \citenamefont
  {Zhu}}]{baeExcitoncoupledCoherentMagnons2022}%
  \BibitemOpen
  \bibfield  {author} {\bibinfo {author} {\bibfnamefont {Y.~J.}\ \bibnamefont
  {Bae}}, \bibinfo {author} {\bibfnamefont {J.}~\bibnamefont {Wang}}, \bibinfo
  {author} {\bibfnamefont {A.}~\bibnamefont {Scheie}}, \bibinfo {author}
  {\bibfnamefont {J.}~\bibnamefont {Xu}}, \bibinfo {author} {\bibfnamefont
  {D.~G.}\ \bibnamefont {Chica}}, \bibinfo {author} {\bibfnamefont {G.~M.}\
  \bibnamefont {Diederich}}, \bibinfo {author} {\bibfnamefont {J.}~\bibnamefont
  {Cenker}}, \bibinfo {author} {\bibfnamefont {M.~E.}\ \bibnamefont {Ziebel}},
  \bibinfo {author} {\bibfnamefont {Y.}~\bibnamefont {Bai}}, \bibinfo {author}
  {\bibfnamefont {H.}~\bibnamefont {Ren}}, \bibinfo {author} {\bibfnamefont
  {C.~R.}\ \bibnamefont {Dean}}, \bibinfo {author} {\bibfnamefont
  {M.}~\bibnamefont {Delor}}, \bibinfo {author} {\bibfnamefont
  {X.}~\bibnamefont {Xu}}, \bibinfo {author} {\bibfnamefont {X.}~\bibnamefont
  {Roy}}, \bibinfo {author} {\bibfnamefont {A.~D.}\ \bibnamefont {Kent}},\ and\
  \bibinfo {author} {\bibfnamefont {X.}~\bibnamefont {Zhu}},\ }\href
  {https://doi.org/10.1038/s41586-022-05024-1} {\bibfield  {journal} {\bibinfo
  {journal} {Nature}\ }\textbf {\bibinfo {volume} {609}},\ \bibinfo {pages}
  {282} (\bibinfo {year} {2022})}\BibitemShut {NoStop}%
\bibitem [{\citenamefont {Scheie}\ \emph {et~al.}(2022)\citenamefont {Scheie},
  \citenamefont {Ziebel}, \citenamefont {Chica}, \citenamefont {Bae},
  \citenamefont {Wang}, \citenamefont {Kolesnikov}, \citenamefont {Zhu},\ and\
  \citenamefont {Roy}}]{scheieSpinWavesMagnetic2022}%
  \BibitemOpen
  \bibfield  {author} {\bibinfo {author} {\bibfnamefont {A.}~\bibnamefont
  {Scheie}}, \bibinfo {author} {\bibfnamefont {M.}~\bibnamefont {Ziebel}},
  \bibinfo {author} {\bibfnamefont {D.~G.}\ \bibnamefont {Chica}}, \bibinfo
  {author} {\bibfnamefont {Y.~J.}\ \bibnamefont {Bae}}, \bibinfo {author}
  {\bibfnamefont {X.}~\bibnamefont {Wang}}, \bibinfo {author} {\bibfnamefont
  {A.~I.}\ \bibnamefont {Kolesnikov}}, \bibinfo {author} {\bibfnamefont
  {X.}~\bibnamefont {Zhu}},\ and\ \bibinfo {author} {\bibfnamefont
  {X.}~\bibnamefont {Roy}},\ }\href {https://doi.org/10.1002/advs.202202467}
  {\bibfield  {journal} {\bibinfo  {journal} {Advanced Science}\ }\textbf
  {\bibinfo {volume} {9}},\ \bibinfo {pages} {2202467} (\bibinfo {year}
  {2022})}\BibitemShut {NoStop}%
\bibitem [{\citenamefont {{Lachance-Quirion}}\ \emph
  {et~al.}(2020)\citenamefont {{Lachance-Quirion}}, \citenamefont {Wolski},
  \citenamefont {Tabuchi}, \citenamefont {Kono}, \citenamefont {Usami},\ and\
  \citenamefont
  {Nakamura}}]{lachance-quirionEntanglementbasedSingleshotDetection2020}%
  \BibitemOpen
  \bibfield  {author} {\bibinfo {author} {\bibfnamefont {D.}~\bibnamefont
  {{Lachance-Quirion}}}, \bibinfo {author} {\bibfnamefont {S.~P.}\ \bibnamefont
  {Wolski}}, \bibinfo {author} {\bibfnamefont {Y.}~\bibnamefont {Tabuchi}},
  \bibinfo {author} {\bibfnamefont {S.}~\bibnamefont {Kono}}, \bibinfo {author}
  {\bibfnamefont {K.}~\bibnamefont {Usami}},\ and\ \bibinfo {author}
  {\bibfnamefont {Y.}~\bibnamefont {Nakamura}},\ }\href
  {https://doi.org/10.1126/science.aaz9236} {\bibfield  {journal} {\bibinfo
  {journal} {Science}\ }\textbf {\bibinfo {volume} {367}},\ \bibinfo {pages}
  {425} (\bibinfo {year} {2020})}\BibitemShut {NoStop}%
\bibitem [{\citenamefont {Pirro}\ \emph {et~al.}(2021)\citenamefont {Pirro},
  \citenamefont {Vasyuchka}, \citenamefont {Serga},\ and\ \citenamefont
  {Hillebrands}}]{pirroAdvancesCoherentMagnonics2021}%
  \BibitemOpen
  \bibfield  {author} {\bibinfo {author} {\bibfnamefont {P.}~\bibnamefont
  {Pirro}}, \bibinfo {author} {\bibfnamefont {V.~I.}\ \bibnamefont
  {Vasyuchka}}, \bibinfo {author} {\bibfnamefont {A.~A.}\ \bibnamefont
  {Serga}},\ and\ \bibinfo {author} {\bibfnamefont {B.}~\bibnamefont
  {Hillebrands}},\ }\href {https://doi.org/10.1038/s41578-021-00332-w}
  {\bibfield  {journal} {\bibinfo  {journal} {Nat Rev Mater}\ }\textbf
  {\bibinfo {volume} {6}},\ \bibinfo {pages} {1114} (\bibinfo {year}
  {2021})}\BibitemShut {NoStop}%
\bibitem [{\citenamefont {Cornelissen}\ \emph {et~al.}(2015)\citenamefont
  {Cornelissen}, \citenamefont {Liu}, \citenamefont {Duine}, \citenamefont
  {Youssef},\ and\ \citenamefont {{van
  Wees}}}]{cornelissenLongdistanceTransportMagnon2015}%
  \BibitemOpen
  \bibfield  {author} {\bibinfo {author} {\bibfnamefont {L.~J.}\ \bibnamefont
  {Cornelissen}}, \bibinfo {author} {\bibfnamefont {J.}~\bibnamefont {Liu}},
  \bibinfo {author} {\bibfnamefont {R.~A.}\ \bibnamefont {Duine}}, \bibinfo
  {author} {\bibfnamefont {J.~B.}\ \bibnamefont {Youssef}},\ and\ \bibinfo
  {author} {\bibfnamefont {B.~J.}\ \bibnamefont {{van Wees}}},\ }\href
  {https://doi.org/10.1038/nphys3465} {\bibfield  {journal} {\bibinfo
  {journal} {Nature Phys}\ }\textbf {\bibinfo {volume} {11}},\ \bibinfo {pages}
  {1022} (\bibinfo {year} {2015})}\BibitemShut {NoStop}%
\bibitem [{\citenamefont {Lebrun}\ \emph {et~al.}(2018)\citenamefont {Lebrun},
  \citenamefont {Ross}, \citenamefont {Bender}, \citenamefont {Qaiumzadeh},
  \citenamefont {Baldrati}, \citenamefont {Cramer}, \citenamefont {Brataas},
  \citenamefont {Duine},\ and\ \citenamefont
  {Kl{\"a}ui}}]{lebrunTunableLongdistanceSpin2018}%
  \BibitemOpen
  \bibfield  {author} {\bibinfo {author} {\bibfnamefont {R.}~\bibnamefont
  {Lebrun}}, \bibinfo {author} {\bibfnamefont {A.}~\bibnamefont {Ross}},
  \bibinfo {author} {\bibfnamefont {S.~A.}\ \bibnamefont {Bender}}, \bibinfo
  {author} {\bibfnamefont {A.}~\bibnamefont {Qaiumzadeh}}, \bibinfo {author}
  {\bibfnamefont {L.}~\bibnamefont {Baldrati}}, \bibinfo {author}
  {\bibfnamefont {J.}~\bibnamefont {Cramer}}, \bibinfo {author} {\bibfnamefont
  {A.}~\bibnamefont {Brataas}}, \bibinfo {author} {\bibfnamefont {R.~A.}\
  \bibnamefont {Duine}},\ and\ \bibinfo {author} {\bibfnamefont
  {M.}~\bibnamefont {Kl{\"a}ui}},\ }\href
  {https://doi.org/10.1038/s41586-018-0490-7} {\bibfield  {journal} {\bibinfo
  {journal} {Nature}\ }\textbf {\bibinfo {volume} {561}},\ \bibinfo {pages}
  {222} (\bibinfo {year} {2018})}\BibitemShut {NoStop}%
\bibitem [{\citenamefont {Lebrun}\ \emph {et~al.}(2020)\citenamefont {Lebrun},
  \citenamefont {Ross}, \citenamefont {Gomonay}, \citenamefont {Baltz},
  \citenamefont {Ebels}, \citenamefont {Barra}, \citenamefont {Qaiumzadeh},
  \citenamefont {Brataas}, \citenamefont {Sinova},\ and\ \citenamefont
  {Kl{\"a}ui}}]{lebrunLongdistanceSpintransportMorin2020}%
  \BibitemOpen
  \bibfield  {author} {\bibinfo {author} {\bibfnamefont {R.}~\bibnamefont
  {Lebrun}}, \bibinfo {author} {\bibfnamefont {A.}~\bibnamefont {Ross}},
  \bibinfo {author} {\bibfnamefont {O.}~\bibnamefont {Gomonay}}, \bibinfo
  {author} {\bibfnamefont {V.}~\bibnamefont {Baltz}}, \bibinfo {author}
  {\bibfnamefont {U.}~\bibnamefont {Ebels}}, \bibinfo {author} {\bibfnamefont
  {A.-L.}\ \bibnamefont {Barra}}, \bibinfo {author} {\bibfnamefont
  {A.}~\bibnamefont {Qaiumzadeh}}, \bibinfo {author} {\bibfnamefont
  {A.}~\bibnamefont {Brataas}}, \bibinfo {author} {\bibfnamefont
  {J.}~\bibnamefont {Sinova}},\ and\ \bibinfo {author} {\bibfnamefont
  {M.}~\bibnamefont {Kl{\"a}ui}},\ }\href
  {https://doi.org/10.1038/s41467-020-20155-7} {\bibfield  {journal} {\bibinfo
  {journal} {Nat Commun}\ }\textbf {\bibinfo {volume} {11}},\ \bibinfo {pages}
  {6332} (\bibinfo {year} {2020})}\BibitemShut {NoStop}%
\bibitem [{\citenamefont {Han}\ \emph {et~al.}(2020)\citenamefont {Han},
  \citenamefont {Zhang}, \citenamefont {Bi}, \citenamefont {Fan}, \citenamefont
  {Safi}, \citenamefont {Xiang}, \citenamefont {Finley}, \citenamefont {Fu},
  \citenamefont {Cheng},\ and\ \citenamefont
  {Liu}}]{hanBirefringencelikeSpinTransport2020}%
  \BibitemOpen
  \bibfield  {author} {\bibinfo {author} {\bibfnamefont {J.}~\bibnamefont
  {Han}}, \bibinfo {author} {\bibfnamefont {P.}~\bibnamefont {Zhang}}, \bibinfo
  {author} {\bibfnamefont {Z.}~\bibnamefont {Bi}}, \bibinfo {author}
  {\bibfnamefont {Y.}~\bibnamefont {Fan}}, \bibinfo {author} {\bibfnamefont
  {T.~S.}\ \bibnamefont {Safi}}, \bibinfo {author} {\bibfnamefont
  {J.}~\bibnamefont {Xiang}}, \bibinfo {author} {\bibfnamefont
  {J.}~\bibnamefont {Finley}}, \bibinfo {author} {\bibfnamefont
  {L.}~\bibnamefont {Fu}}, \bibinfo {author} {\bibfnamefont {R.}~\bibnamefont
  {Cheng}},\ and\ \bibinfo {author} {\bibfnamefont {L.}~\bibnamefont {Liu}},\
  }\href {https://doi.org/10.1038/s41565-020-0703-8} {\bibfield  {journal}
  {\bibinfo  {journal} {Nature Nanotechnology}\ }\textbf {\bibinfo {volume}
  {15}},\ \bibinfo {pages} {563} (\bibinfo {year} {2020})}\BibitemShut
  {NoStop}%
\bibitem [{\citenamefont {Wei}\ \emph {et~al.}(2022)\citenamefont {Wei},
  \citenamefont {Santos}, \citenamefont {Lusero}, \citenamefont {Bauer},
  \citenamefont {Ben~Youssef},\ and\ \citenamefont {{van
  Wees}}}]{weiGiantMagnonSpin2022}%
  \BibitemOpen
  \bibfield  {author} {\bibinfo {author} {\bibfnamefont {X.-Y.}\ \bibnamefont
  {Wei}}, \bibinfo {author} {\bibfnamefont {O.~A.}\ \bibnamefont {Santos}},
  \bibinfo {author} {\bibfnamefont {C.~H.~S.}\ \bibnamefont {Lusero}}, \bibinfo
  {author} {\bibfnamefont {G.~E.~W.}\ \bibnamefont {Bauer}}, \bibinfo {author}
  {\bibfnamefont {J.}~\bibnamefont {Ben~Youssef}},\ and\ \bibinfo {author}
  {\bibfnamefont {B.~J.}\ \bibnamefont {{van Wees}}},\ }\href
  {https://doi.org/10.1038/s41563-022-01369-0} {\bibfield  {journal} {\bibinfo
  {journal} {Nat. Mater.}\ }\textbf {\bibinfo {volume} {9}},\ \bibinfo {pages}
  {1352} (\bibinfo {year} {2022})}\BibitemShut {NoStop}%
\bibitem [{\citenamefont {Jungwirth}\ \emph {et~al.}(2016)\citenamefont
  {Jungwirth}, \citenamefont {Marti}, \citenamefont {Wadley},\ and\
  \citenamefont {Wunderlich}}]{jungwirthAntiferromagneticSpintronics2016}%
  \BibitemOpen
  \bibfield  {author} {\bibinfo {author} {\bibfnamefont {T.}~\bibnamefont
  {Jungwirth}}, \bibinfo {author} {\bibfnamefont {X.}~\bibnamefont {Marti}},
  \bibinfo {author} {\bibfnamefont {P.}~\bibnamefont {Wadley}},\ and\ \bibinfo
  {author} {\bibfnamefont {J.}~\bibnamefont {Wunderlich}},\ }\href
  {https://doi.org/10.1038/nnano.2016.18} {\bibfield  {journal} {\bibinfo
  {journal} {Nature Nanotechnology}\ }\textbf {\bibinfo {volume} {11}},\
  \bibinfo {pages} {231} (\bibinfo {year} {2016})}\BibitemShut {NoStop}%
\bibitem [{\citenamefont {Baltz}\ \emph {et~al.}(2018)\citenamefont {Baltz},
  \citenamefont {Manchon}, \citenamefont {Tsoi}, \citenamefont {Moriyama},
  \citenamefont {Ono},\ and\ \citenamefont
  {Tserkovnyak}}]{baltzAntiferromagneticSpintronics2018}%
  \BibitemOpen
  \bibfield  {author} {\bibinfo {author} {\bibfnamefont {V.}~\bibnamefont
  {Baltz}}, \bibinfo {author} {\bibfnamefont {A.}~\bibnamefont {Manchon}},
  \bibinfo {author} {\bibfnamefont {M.}~\bibnamefont {Tsoi}}, \bibinfo {author}
  {\bibfnamefont {T.}~\bibnamefont {Moriyama}}, \bibinfo {author}
  {\bibfnamefont {T.}~\bibnamefont {Ono}},\ and\ \bibinfo {author}
  {\bibfnamefont {Y.}~\bibnamefont {Tserkovnyak}},\ }\href
  {https://doi.org/10.1103/RevModPhys.90.015005} {\bibfield  {journal}
  {\bibinfo  {journal} {Rev. Mod. Phys.}\ }\textbf {\bibinfo {volume} {90}},\
  \bibinfo {pages} {015005} (\bibinfo {year} {2018})}\BibitemShut {NoStop}%
\bibitem [{\citenamefont {Xing}\ \emph {et~al.}(2019)\citenamefont {Xing},
  \citenamefont {Qiu}, \citenamefont {Wang}, \citenamefont {Yao}, \citenamefont
  {Ma}, \citenamefont {Cai}, \citenamefont {Jia}, \citenamefont {Xie},\ and\
  \citenamefont {Han}}]{xingMagnonTransportQuasiTwoDimensional2019}%
  \BibitemOpen
  \bibfield  {author} {\bibinfo {author} {\bibfnamefont {W.}~\bibnamefont
  {Xing}}, \bibinfo {author} {\bibfnamefont {L.}~\bibnamefont {Qiu}}, \bibinfo
  {author} {\bibfnamefont {X.}~\bibnamefont {Wang}}, \bibinfo {author}
  {\bibfnamefont {Y.}~\bibnamefont {Yao}}, \bibinfo {author} {\bibfnamefont
  {Y.}~\bibnamefont {Ma}}, \bibinfo {author} {\bibfnamefont {R.}~\bibnamefont
  {Cai}}, \bibinfo {author} {\bibfnamefont {S.}~\bibnamefont {Jia}}, \bibinfo
  {author} {\bibfnamefont {X.~C.}\ \bibnamefont {Xie}},\ and\ \bibinfo {author}
  {\bibfnamefont {W.}~\bibnamefont {Han}},\ }\href
  {https://doi.org/10.1103/PhysRevX.9.011026} {\bibfield  {journal} {\bibinfo
  {journal} {Phys. Rev. X}\ }\textbf {\bibinfo {volume} {9}},\ \bibinfo {pages}
  {011026} (\bibinfo {year} {2019})}\BibitemShut {NoStop}%
\bibitem [{\citenamefont {Hoogeboom}\ and\ \citenamefont {{van
  Wees}}(2020)}]{hoogeboomNonlocalSpinSeebeck2020}%
  \BibitemOpen
  \bibfield  {author} {\bibinfo {author} {\bibfnamefont {G.~R.}\ \bibnamefont
  {Hoogeboom}}\ and\ \bibinfo {author} {\bibfnamefont {B.~J.}\ \bibnamefont
  {{van Wees}}},\ }\href {https://doi.org/10.1103/PhysRevB.102.214415}
  {\bibfield  {journal} {\bibinfo  {journal} {Phys. Rev. B}\ }\textbf {\bibinfo
  {volume} {102}},\ \bibinfo {pages} {214415} (\bibinfo {year}
  {2020})}\BibitemShut {NoStop}%
\bibitem [{\citenamefont {Kalashnikova}\ \emph {et~al.}(2007)\citenamefont
  {Kalashnikova}, \citenamefont {Kimel}, \citenamefont {Pisarev}, \citenamefont
  {Gridnev}, \citenamefont {Kirilyuk},\ and\ \citenamefont
  {Rasing}}]{kalashnikovaImpulsiveGenerationCoherent2007}%
  \BibitemOpen
  \bibfield  {author} {\bibinfo {author} {\bibfnamefont {A.~M.}\ \bibnamefont
  {Kalashnikova}}, \bibinfo {author} {\bibfnamefont {A.~V.}\ \bibnamefont
  {Kimel}}, \bibinfo {author} {\bibfnamefont {R.~V.}\ \bibnamefont {Pisarev}},
  \bibinfo {author} {\bibfnamefont {V.~N.}\ \bibnamefont {Gridnev}}, \bibinfo
  {author} {\bibfnamefont {A.}~\bibnamefont {Kirilyuk}},\ and\ \bibinfo
  {author} {\bibfnamefont {T.}~\bibnamefont {Rasing}},\ }\href
  {https://doi.org/10.1103/PhysRevLett.99.167205} {\bibfield  {journal}
  {\bibinfo  {journal} {Phys. Rev. Lett.}\ }\textbf {\bibinfo {volume} {99}},\
  \bibinfo {pages} {167205} (\bibinfo {year} {2007})}\BibitemShut {NoStop}%
\bibitem [{\citenamefont {Satoh}\ \emph {et~al.}(2010)\citenamefont {Satoh},
  \citenamefont {Cho}, \citenamefont {Iida}, \citenamefont {Shimura},
  \citenamefont {Kuroda}, \citenamefont {Ueda}, \citenamefont {Ueda},
  \citenamefont {Ivanov}, \citenamefont {Nori},\ and\ \citenamefont
  {Fiebig}}]{satohSpinOscillationsAntiferromagnetic2010}%
  \BibitemOpen
  \bibfield  {author} {\bibinfo {author} {\bibfnamefont {T.}~\bibnamefont
  {Satoh}}, \bibinfo {author} {\bibfnamefont {S.-J.}\ \bibnamefont {Cho}},
  \bibinfo {author} {\bibfnamefont {R.}~\bibnamefont {Iida}}, \bibinfo {author}
  {\bibfnamefont {T.}~\bibnamefont {Shimura}}, \bibinfo {author} {\bibfnamefont
  {K.}~\bibnamefont {Kuroda}}, \bibinfo {author} {\bibfnamefont
  {H.}~\bibnamefont {Ueda}}, \bibinfo {author} {\bibfnamefont {Y.}~\bibnamefont
  {Ueda}}, \bibinfo {author} {\bibfnamefont {B.~A.}\ \bibnamefont {Ivanov}},
  \bibinfo {author} {\bibfnamefont {F.}~\bibnamefont {Nori}},\ and\ \bibinfo
  {author} {\bibfnamefont {M.}~\bibnamefont {Fiebig}},\ }\href
  {https://doi.org/10.1103/PhysRevLett.105.077402} {\bibfield  {journal}
  {\bibinfo  {journal} {Phys. Rev. Lett.}\ }\textbf {\bibinfo {volume} {105}},\
  \bibinfo {pages} {077402} (\bibinfo {year} {2010})}\BibitemShut {NoStop}%
\bibitem [{\citenamefont {Tzschaschel}\ \emph {et~al.}(2017)\citenamefont
  {Tzschaschel}, \citenamefont {Otani}, \citenamefont {Iida}, \citenamefont
  {Shimura}, \citenamefont {Ueda}, \citenamefont {G{\"u}nther}, \citenamefont
  {Fiebig},\ and\ \citenamefont
  {Satoh}}]{tzschaschelUltrafastOpticalExcitation2017}%
  \BibitemOpen
  \bibfield  {author} {\bibinfo {author} {\bibfnamefont {C.}~\bibnamefont
  {Tzschaschel}}, \bibinfo {author} {\bibfnamefont {K.}~\bibnamefont {Otani}},
  \bibinfo {author} {\bibfnamefont {R.}~\bibnamefont {Iida}}, \bibinfo {author}
  {\bibfnamefont {T.}~\bibnamefont {Shimura}}, \bibinfo {author} {\bibfnamefont
  {H.}~\bibnamefont {Ueda}}, \bibinfo {author} {\bibfnamefont {S.}~\bibnamefont
  {G{\"u}nther}}, \bibinfo {author} {\bibfnamefont {M.}~\bibnamefont
  {Fiebig}},\ and\ \bibinfo {author} {\bibfnamefont {T.}~\bibnamefont
  {Satoh}},\ }\href {https://doi.org/10.1103/PhysRevB.95.174407} {\bibfield
  {journal} {\bibinfo  {journal} {Phys. Rev. B}\ }\textbf {\bibinfo {volume}
  {95}},\ \bibinfo {pages} {174407} (\bibinfo {year} {2017})}\BibitemShut
  {NoStop}%
\bibitem [{\citenamefont {Sonin}(2010)}]{soninSpinCurrentsSpin2010}%
  \BibitemOpen
  \bibfield  {author} {\bibinfo {author} {\bibfnamefont {E.}~\bibnamefont
  {Sonin}},\ }\href {https://doi.org/10.1080/00018731003739943} {\bibfield
  {journal} {\bibinfo  {journal} {Advances in Physics}\ }\textbf {\bibinfo
  {volume} {59}},\ \bibinfo {pages} {181} (\bibinfo {year} {2010})}\BibitemShut
  {NoStop}%
\bibitem [{\citenamefont {Sonin}(2020)}]{soninSuperfluidSpinTransport2020}%
  \BibitemOpen
  \bibfield  {author} {\bibinfo {author} {\bibfnamefont {E.~B.}\ \bibnamefont
  {Sonin}},\ }\href {https://doi.org/10.1063/10.0001046} {\bibfield  {journal}
  {\bibinfo  {journal} {Low Temperature Physics}\ }\textbf {\bibinfo {volume}
  {46}},\ \bibinfo {pages} {436} (\bibinfo {year} {2020})}\BibitemShut
  {NoStop}%
\bibitem [{\citenamefont {Diederich}\ \emph {et~al.}(2023)\citenamefont
  {Diederich}, \citenamefont {Cenker}, \citenamefont {Ren}, \citenamefont
  {Fonseca}, \citenamefont {Chica}, \citenamefont {Bae}, \citenamefont {Zhu},
  \citenamefont {Roy}, \citenamefont {Cao}, \citenamefont {Xiao},\ and\
  \citenamefont {Xu}}]{diederichTunableInteractionExcitons2023}%
  \BibitemOpen
  \bibfield  {author} {\bibinfo {author} {\bibfnamefont {G.~M.}\ \bibnamefont
  {Diederich}}, \bibinfo {author} {\bibfnamefont {J.}~\bibnamefont {Cenker}},
  \bibinfo {author} {\bibfnamefont {Y.}~\bibnamefont {Ren}}, \bibinfo {author}
  {\bibfnamefont {J.}~\bibnamefont {Fonseca}}, \bibinfo {author} {\bibfnamefont
  {D.~G.}\ \bibnamefont {Chica}}, \bibinfo {author} {\bibfnamefont {Y.~J.}\
  \bibnamefont {Bae}}, \bibinfo {author} {\bibfnamefont {X.}~\bibnamefont
  {Zhu}}, \bibinfo {author} {\bibfnamefont {X.}~\bibnamefont {Roy}}, \bibinfo
  {author} {\bibfnamefont {T.}~\bibnamefont {Cao}}, \bibinfo {author}
  {\bibfnamefont {D.}~\bibnamefont {Xiao}},\ and\ \bibinfo {author}
  {\bibfnamefont {X.}~\bibnamefont {Xu}},\ }\href
  {https://doi.org/10.1038/s41565-022-01259-1} {\bibfield  {journal} {\bibinfo
  {journal} {Nature Nanotechnology}\ }\textbf {\bibinfo {volume} {18}},\
  \bibinfo {pages} {23} (\bibinfo {year} {2023})}\BibitemShut {NoStop}%
\bibitem [{\citenamefont {Damon}\ and\ \citenamefont
  {Eshbach}(1961)}]{damonMagnetostaticModesFerromagnet1961}%
  \BibitemOpen
  \bibfield  {author} {\bibinfo {author} {\bibfnamefont {R.}~\bibnamefont
  {Damon}}\ and\ \bibinfo {author} {\bibfnamefont {J.}~\bibnamefont
  {Eshbach}},\ }\href {https://doi.org/10.1016/0022-3697(61)90041-5} {\bibfield
   {journal} {\bibinfo  {journal} {Journal of Physics and Chemistry of Solids}\
  }\textbf {\bibinfo {volume} {19}},\ \bibinfo {pages} {308} (\bibinfo {year}
  {1961})}\BibitemShut {NoStop}%
\bibitem [{\citenamefont {Hurben}\ and\ \citenamefont
  {Patton}(1996)}]{hurbenTheoryMagnetostaticWaves1996}%
  \BibitemOpen
  \bibfield  {author} {\bibinfo {author} {\bibfnamefont {M.}~\bibnamefont
  {Hurben}}\ and\ \bibinfo {author} {\bibfnamefont {C.}~\bibnamefont
  {Patton}},\ }\href {https://doi.org/10.1016/S0304-8853(96)00294-6} {\bibfield
   {journal} {\bibinfo  {journal} {Journal of Magnetism and Magnetic
  Materials}\ }\textbf {\bibinfo {volume} {163}},\ \bibinfo {pages} {39}
  (\bibinfo {year} {1996})}\BibitemShut {NoStop}%
\bibitem [{\citenamefont {Satoh}\ \emph {et~al.}(2012)\citenamefont {Satoh},
  \citenamefont {Terui}, \citenamefont {Moriya}, \citenamefont {Ivanov},
  \citenamefont {Ando}, \citenamefont {Saitoh}, \citenamefont {Shimura},\ and\
  \citenamefont {Kuroda}}]{satohDirectionalControlSpinwave2012}%
  \BibitemOpen
  \bibfield  {author} {\bibinfo {author} {\bibfnamefont {T.}~\bibnamefont
  {Satoh}}, \bibinfo {author} {\bibfnamefont {Y.}~\bibnamefont {Terui}},
  \bibinfo {author} {\bibfnamefont {R.}~\bibnamefont {Moriya}}, \bibinfo
  {author} {\bibfnamefont {B.~A.}\ \bibnamefont {Ivanov}}, \bibinfo {author}
  {\bibfnamefont {K.}~\bibnamefont {Ando}}, \bibinfo {author} {\bibfnamefont
  {E.}~\bibnamefont {Saitoh}}, \bibinfo {author} {\bibfnamefont
  {T.}~\bibnamefont {Shimura}},\ and\ \bibinfo {author} {\bibfnamefont
  {K.}~\bibnamefont {Kuroda}},\ }\href
  {https://doi.org/10.1038/nphoton.2012.218} {\bibfield  {journal} {\bibinfo
  {journal} {Nature Photon}\ }\textbf {\bibinfo {volume} {6}},\ \bibinfo
  {pages} {662} (\bibinfo {year} {2012})}\BibitemShut {NoStop}%
\bibitem [{\citenamefont {Demokritov}\ \emph {et~al.}(2006)\citenamefont
  {Demokritov}, \citenamefont {Demidov}, \citenamefont {Dzyapko}, \citenamefont
  {Melkov}, \citenamefont {Serga}, \citenamefont {Hillebrands},\ and\
  \citenamefont {Slavin}}]{demokritovBoseEinsteinCondensation2006}%
  \BibitemOpen
  \bibfield  {author} {\bibinfo {author} {\bibfnamefont {S.~O.}\ \bibnamefont
  {Demokritov}}, \bibinfo {author} {\bibfnamefont {V.~E.}\ \bibnamefont
  {Demidov}}, \bibinfo {author} {\bibfnamefont {O.}~\bibnamefont {Dzyapko}},
  \bibinfo {author} {\bibfnamefont {G.~A.}\ \bibnamefont {Melkov}}, \bibinfo
  {author} {\bibfnamefont {A.~A.}\ \bibnamefont {Serga}}, \bibinfo {author}
  {\bibfnamefont {B.}~\bibnamefont {Hillebrands}},\ and\ \bibinfo {author}
  {\bibfnamefont {A.~N.}\ \bibnamefont {Slavin}},\ }\href
  {https://doi.org/10.1038/nature05117} {\bibfield  {journal} {\bibinfo
  {journal} {Nature}\ }\textbf {\bibinfo {volume} {443}},\ \bibinfo {pages}
  {430} (\bibinfo {year} {2006})}\BibitemShut {NoStop}%
\bibitem [{\citenamefont {Matsumoto}\ and\ \citenamefont
  {Murakami}(2011)}]{matsumotoTheoreticalPredictionRotating2011}%
  \BibitemOpen
  \bibfield  {author} {\bibinfo {author} {\bibfnamefont {R.}~\bibnamefont
  {Matsumoto}}\ and\ \bibinfo {author} {\bibfnamefont {S.}~\bibnamefont
  {Murakami}},\ }\href {https://doi.org/10.1103/PhysRevLett.106.197202}
  {\bibfield  {journal} {\bibinfo  {journal} {Phys. Rev. Lett.}\ }\textbf
  {\bibinfo {volume} {106}},\ \bibinfo {pages} {197202} (\bibinfo {year}
  {2011})}\BibitemShut {NoStop}%
\bibitem [{\citenamefont {Camley}(1980)}]{camleyLongWavelengthSurfaceSpin1980}%
  \BibitemOpen
  \bibfield  {author} {\bibinfo {author} {\bibfnamefont {R.~E.}\ \bibnamefont
  {Camley}},\ }\href {https://doi.org/10.1103/PhysRevLett.45.283} {\bibfield
  {journal} {\bibinfo  {journal} {Phys. Rev. Lett.}\ }\textbf {\bibinfo
  {volume} {45}},\ \bibinfo {pages} {283} (\bibinfo {year} {1980})}\BibitemShut
  {NoStop}%
\bibitem [{\citenamefont {L{\"u}thi}\ \emph {et~al.}(1983)\citenamefont
  {L{\"u}thi}, \citenamefont {Mills},\ and\ \citenamefont
  {Camley}}]{luthiSurfaceSpinWaves1983}%
  \BibitemOpen
  \bibfield  {author} {\bibinfo {author} {\bibfnamefont {B.}~\bibnamefont
  {L{\"u}thi}}, \bibinfo {author} {\bibfnamefont {D.~L.}\ \bibnamefont
  {Mills}},\ and\ \bibinfo {author} {\bibfnamefont {R.~E.}\ \bibnamefont
  {Camley}},\ }\href {https://doi.org/10.1103/PhysRevB.28.1475} {\bibfield
  {journal} {\bibinfo  {journal} {Physical Review B}\ }\textbf {\bibinfo
  {volume} {28}},\ \bibinfo {pages} {1475} (\bibinfo {year}
  {1983})}\BibitemShut {NoStop}%
\bibitem [{\citenamefont {Stamps}\ and\ \citenamefont
  {Camley}(1986)}]{stampsMagnetostaticModesThin1986}%
  \BibitemOpen
  \bibfield  {author} {\bibinfo {author} {\bibfnamefont {R.}~\bibnamefont
  {Stamps}}\ and\ \bibinfo {author} {\bibfnamefont {R.}~\bibnamefont
  {Camley}},\ }\href {https://doi.org/10.1016/0304-8853(86)90261-1} {\bibfield
  {journal} {\bibinfo  {journal} {Journal of Magnetism and Magnetic Materials}\
  }\textbf {\bibinfo {volume} {54--57}},\ \bibinfo {pages} {803} (\bibinfo
  {year} {1986})}\BibitemShut {NoStop}%
\bibitem [{\citenamefont {Lee}\ \emph {et~al.}(2023)\citenamefont {Lee},
  \citenamefont {Sun}, \citenamefont {Ye}, \citenamefont {Rathi}, \citenamefont
  {Wang}, \citenamefont {Lu}, \citenamefont {Moore}, \citenamefont
  {Checkelsky},\ and\ \citenamefont
  {Orenstein}}]{leeSpinWavepacketsKagome2023}%
  \BibitemOpen
  \bibfield  {author} {\bibinfo {author} {\bibfnamefont {C.}~\bibnamefont
  {Lee}}, \bibinfo {author} {\bibfnamefont {Y.}~\bibnamefont {Sun}}, \bibinfo
  {author} {\bibfnamefont {L.}~\bibnamefont {Ye}}, \bibinfo {author}
  {\bibfnamefont {S.}~\bibnamefont {Rathi}}, \bibinfo {author} {\bibfnamefont
  {K.}~\bibnamefont {Wang}}, \bibinfo {author} {\bibfnamefont {Y.-M.}\
  \bibnamefont {Lu}}, \bibinfo {author} {\bibfnamefont {J.}~\bibnamefont
  {Moore}}, \bibinfo {author} {\bibfnamefont {J.~G.}\ \bibnamefont
  {Checkelsky}},\ and\ \bibinfo {author} {\bibfnamefont {J.}~\bibnamefont
  {Orenstein}},\ }\href {https://doi.org/10.1073/pnas.2220589120} {\bibfield
  {journal} {\bibinfo  {journal} {Proceedings of the National Academy of
  Sciences}\ }\textbf {\bibinfo {volume} {120}},\ \bibinfo {pages}
  {e2220589120} (\bibinfo {year} {2023})}\BibitemShut {NoStop}%
\end{thebibliography}%

\end{document}